\documentclass[aps,prb,floatfix,twocolumn,showpacs,reprint,superscriptaddress]{revtex4-1}
\usepackage{amsfonts}
\usepackage{mathrsfs}
\usepackage{amsmath}% needed for subequations
\usepackage{color}
\usepackage{graphicx}
\usepackage{bm}% bold maths
\usepackage{amssymb}
\usepackage{xspace}
\usepackage{epstopdf}
\usepackage{dcolumn}% Align table columns on decimal point
\usepackage{longtable}
\usepackage{multirow}
\usepackage[colorlinks=true, letterpaper=true, pdfstartview=FitV, linkcolor=blue, citecolor=blue, urlcolor=blue]{hyperref}

\makeatletter

\newcommand{\Rmnum}[1]{\expandafter\@slowromancap\romannumeral #1@}
\makeatother
\begin{document}
\title{Perfect Valley Filter in Topological Domain Wall}

\author{Hui Pan}
\affiliation{Department of Physics, Beihang University, Beijing 100191, China}

\author{Xin Li}
\affiliation{Department of Physics, Beihang University, Beijing 100191, China}

\author{Fan Zhang}
\email{zhang@utdallas.edu}
\affiliation{Department of Physics, University of Texas at Dallas, Richardson, Texas 75080, USA}

\author{Shengyuan A. Yang}
\email{shengyuan\_yang@sutd.edu.sg}
\affiliation{Engineering Product Development, Singapore University of Technology and Design, Singapore 487372, Singapore}

\begin{abstract}
We propose a realization of perfect valley filters based on the chiral domain-wall channels between a quantum anomalous Hall insulator and a quantum valley Hall insulator. Uniquely, all these channels reside in the same valley and propagate unidirectionally, $100\%$ valley-polarizing passing-by carriers without backscattering. The valley index, the chirality, and the number of the channels are protected by topological charges, controllable by external fields, and detectable by circular dichroism.
\end{abstract}


\maketitle

Many crystalline materials have two or more degenerate energy extrema in momentum space,
which endows the low-energy carriers with a valley degree of freedom.
Following the extensive studies in spintronics,
there has been a growing interest in exploring the possibility of manipulating valley degrees for information processing,
leading to a concept of valleytronics.~\cite{guna2006,ryce2007,xiao2007,yao2008,zhan2011,zhu2012,xiao2012,cai2013}
For practical valleytronics applications,
it is critical to have a controllable and efficient way to generate valley polarization of carriers.
Although recent experiments have succeeded in generating and modulating the valley polarization dynamically by circularly-polarized lights
in monolayer transition metal dichalcogenides,~\cite{zeng2012,mak2012,cao2012,xu2014}
it is yet desired to have valley control by static means in transport.~\cite{Mak-14,Geim-14,Baug14}
To this end, a valley filter is an indispensable component in valleytronics devices.
Several schemes of valley filters have been suggested in the context of graphene,
e.g., by using nano-constriction,~\cite{ryce2007} line defects,~\cite{gunl2011} strain engineering,~\cite{jian2013} or the recently proposed valley-polarized quantum anomalous Hall phase.~\cite{PanH,zhan2011,Pan2014}
Since in these filters propagating modes of both valleys generally co-exist,
it is arduous to achieve perfect filtering due to intervalley scattering caused by lattice-scale defects,
which are challenging to completely eliminate during fabrication.

Here we propose a general scheme for creating a perfect valley filter in honeycomb-lattice materials
utilizing a domain wall (DW)~\cite{mart2008,seme2008,yao2009,zhang1,mele-13,Alden,Vaezi,kim2014,zhang2}
between two topologically-distinct insulating regions:
a quantum anomalous Hall (QAH) insulator~\cite{hald1988,qiao2010,zhan2011} breaking time-reversal symmetry ($\mathcal{T}$)
and a quantum valley Hall (QVH) insulator~\cite{xiao2007,zhan2011} breaking inversion symmetry ($\mathcal{P}$).
Since both symmetries are destroyed, the valley filtering becomes efficient.
It turns out that subgap-propagating chiral channels appear in the DW,
which are dictated by the change of bulk topological charge across the DW.~\cite{volo2003,mart2008,mele-13}
Remarkably enough, all the unidirectional channels reside in the same valley
when the valley-projected topological charge only changes at one designed valley across the DW.
As a result, carriers in the filter could achieve a $100\%$ valley polarization without any backscattering.
Due to its topological origin, the performance of such a filter is robust against scattering and structural imperfections.
We show that both the valley index and the chirality of our proposed elementary filter can be easily controlled by external fields
and can be probed by optical circular dichroism,
which allows versatile valleytronics functionalities achieved in a relatively simple design.

\begin{figure}
\scalebox{0.51}{\includegraphics*{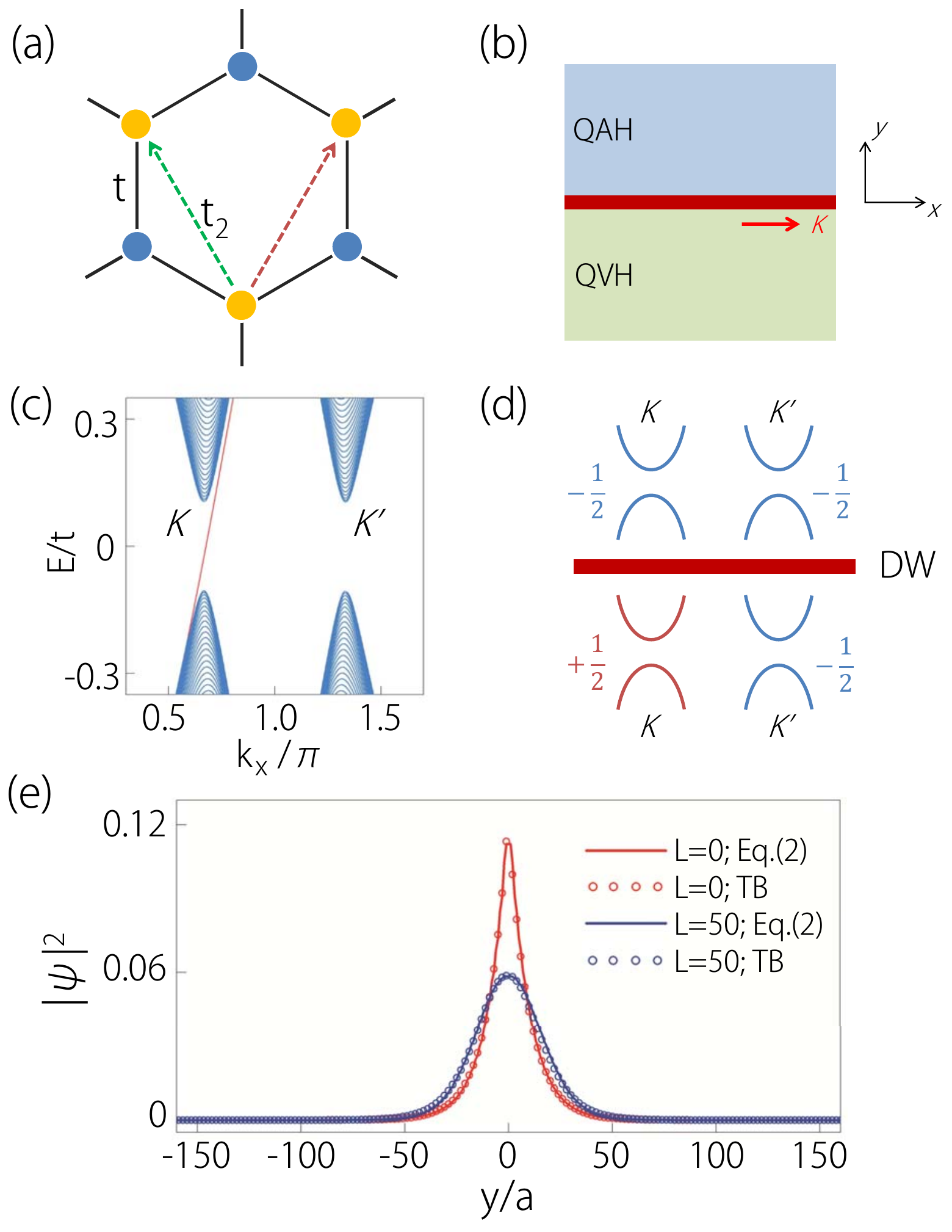}}
\caption{The realization of a perfect valley filter.
(a) The lattice model described by Eq.~(\ref{toyH}). Blue and yellow dots denote the two sublattices.
Representative next-nearest neighbor hopping are indicated by the red (green) arrows with $\nu_{ij}=+\,(-)$.
(b) Schematics of the topological DW proposed in the main text.
(c) The energy spectrum~\cite{param_fig1}  of (b) with a valley-filtered chiral DW channel.
(d) The valley-projected topological charges on the two sides of the DW. (e) Wavefunction distribution for the zero energy mode at valley $K$. Here we take two different DW widths $L$ (in units of $a$) for a DW with a $\tanh(y/L)$ profile for the mass term. The solid lines are from Eq.(\ref{DWmode}) and the circles are solutions from the tight-binding model.}
\label{fig1}
\end{figure}

{\color{blue}{\it Toy model.}}---We first present our basic idea through the analysis of a simple toy model.
Consider a spinless fermion model defined on a 2D honeycomb lattice (see Fig.~\ref{fig1}(a))
\begin{equation}\label{toyH}
\mathcal{H}=t\sum_{\langle i,j\rangle}c_i^\dagger c_j
+it_2\sum_{\langle\langle i,j\rangle\rangle}\nu_{ij}c_i^\dagger c_j+\lambda_v\sum_i\xi_i c_i^\dagger c_i.
\end{equation}
Here the first term represents the nearest neighbor hopping.
The second term, first introduced by Haldane~\cite{hald1988} connects the next-nearest neighbors,
with $\nu_{ij}=+\,(-)$ if the electron makes a left (right) turn when hopping from site $j$ to site $i$.~\cite{kane2005}
The last term is a staggered potential with opposite signs ($\xi_i=\pm 1$) on different sublattices.~\cite{kane2005}
In the absence of the last two terms in~(\ref{toyH}), it is known that near half filling the energy dispersion
is gapless and linear down to the $K$ and $K'$ Dirac points, i.e., the two inequivalent corners of the hexagonal Brillouin zone.
The two valleys are related by $\mathcal{T}$ and $\mathcal{P}$ symmetry operations.
The gapless Dirac point at each valley is protected by a sublattice symmetry and the winding number of an infinitesimal Fermi circle.
Both the Haldane term and the staggered potential open energy gaps at the Dirac points, as each breaks the sublattice symmetry.
Importantly, the Haldane term breaks $\mathcal{T}$ and the staggered potential breaks $\mathcal{P}$.
We consider a DW formed by a spatial separation of the two topologically-distinct states,
e.g., $t_2$ ($\lambda_v$) to be nonzero only for region $y>0$ ($y<0$), as shown in Fig.~\ref{fig1}(b).
Figure~\ref{fig1}(c) plots the energy spectrum with such a DW along the zigzag direction in order for the two valleys to be distinguished.
In the bulk band gap, remarkably, there appears a chiral gapless channel attached to valley $K$,
localized at the DW, and propagating along $+\hat x$ direction.

To gain a better understanding of this chiral gapless valley-filtered channel, we focus on the low-energy physics of the toy model.
Away from the DW, taking $t$ as the largest energy scale, we expand the Hamiltonian around $K$ ($K'$) point and obtain a continuum model $H_\text{eff}=H_0+H_\Delta$,
where the band term $H_0=v_F(\tau_zk_x\sigma_x+k_y\sigma_y)$
and the mass term $H_\Delta$ is $-\lambda_a\tau_z\sigma_z$ ($\lambda_a=3\sqrt{3}t_2$)
for $y\rightarrow+\infty$ and $\lambda_v\sigma_z$ for $y\rightarrow-\infty$.
Here the wave vector $k$ is measured from the $K$ ($K'$) point,
$v_F=3at/2$ is the Fermi velocity with $a$ the distance between nearest neighbors,
$\tau_z=\pm$ denotes the two valleys, Pauli matrices $\bm\sigma$ act on the sublattice pseudospin,
and we assume that $\lambda_v,\lambda_a,v_F>0$.
One observes that $H_\text{eff}$ of valley $K$ is a gapped two-band model with a mass sign reversal across the DW.
It is quintessential that such a Jackiw-Rebbi problem~\cite{jack1976} bears a zero mode:
\begin{equation}\label{DWmode}
\psi_{k_x}(\bm r)=\frac{1}{{A}}\left(
                  \begin{array}{c}
                    1 \\
                    1 \\
                  \end{array}
                \right)\exp\left[ik_x x+\frac{1}{v_F}\int_0^y \Delta(y')dy'\right],
\end{equation}
where ${A}$ is a normalization constant and the mass $\Delta$ is defined through $H_\Delta=\Delta(y)\sigma_z$.
This mode is exponentially localized at the DW, with a chiral dispersion $E=v_F k_x$ around zero energy. In Fig.~\ref{fig1}(e), we plot the wavefunction distribution of the DW zero energy mode for a sharp DW and for a DW with finite width.
Beneficially, the presence of such a chiral gapless DW channel depends only on the designed signs of $\Delta$ at $y=\pm \infty$,
regardless of the detailed DW characteristics, which reflects its topological nature.

The robustness of the chiral gapless DW channel
is intimately connected to the change
of bulk topological charges across the DW.~\cite{volo2003,mart2008,zhan2011,mele-13}
Either side of the DW has a valley-projected topological charge
$\mathcal{N}=\frac{1}{2\pi}\int d^2\bm k \mathit \Omega(\bm k)$,~\cite{mart2008,yao2009}
where $\bm{\mathit \Omega}(\bm k)\equiv\nabla_{\bm k}\times \bm{\mathcal{A}}(\bm k)$ and
$\bm{\mathcal{A}}(\bm k)\equiv\langle u_{\bm k}|i\nabla_{\bm k}|u_{\bm k}\rangle$ are the Berry curvature and the Berry connection of the valence bands with $|u_{\bm k}\rangle$ being the periodic part of the Bloch eigenstate.
The Berry curvature is out-of-plane for the 2D case and concentrated around each valley center in our model.
For the continuum model $H_\text{eff}$, we obtain $\mathcal{N}=\tau_z\text{sgn}(\Delta)/2$.
Since $\Delta=-\lambda_a\tau_z$ in the QAH domain ($y>0$), $\mathcal{N}=-1/2$ for both valleys,
where the valley independence is dictated by $\mathcal{P}$.
Evidently, the total topological charge of both valleys (Chern number) is one, featuring a QAH state.
In the QVH domain with $\Delta=\lambda_v$ ($y<0$), $\mathcal{N}=\tau_z/2$,
where the opposition of topological charges of different valleys is a consequence of $\mathcal{T}$ (in this sense the QVH state is analogous to the quantum spin Hall state~\cite{kane2005,zhan2011}).
These topological charges are illustrated in Fig.~\ref{fig1}(d).
The number of the valley-dependent gapless chiral DW modes is related to the difference of topological charges across the DW by an index theorem.\cite{volo2003,mart2008,seme2008} More specifically, for each valley, $\nu$, the number of chiral modes moving in $+\hat{x}$ direction minus the number of chiral modes moving in $-\hat{x}$ direction, is equal to the difference of the bulk topological charges $\mathcal{N}$ on the two sides of the DW, $\nu=\mathcal{N}(y<0)-\mathcal{N}(y>0)$. It follows that $\nu=1$ for valley $K$ and $\nu=0$ for valley $K'$ in our case.
As a consequence, one chiral gapless channel is protected at valley $K$ whereas none is at valley $K'$.
This topological argument is consistent with our pervious analysis.

The above analysis demonstrates that when across the DW the valley-projected topological charge changes (by an integer) for only one valley,
the DW is guaranteed to possess valley-filtered chiral modes.
Because all the propagating channels in the bulk gap are of the same valley, scattering cannot change the carriers' valley index;
and because of the chiral nature of these channels, there is no backscattering near the Fermi energy (assumed in the bulk gap).
As a result, such a DW is perfectly transparent for one valley but completely opaque for the other valley,
hence behaving as a perfect valley filter.
Regarding the valley-filter operation, the DW is hardly necessary to be exactly along the zigzag direction,
as long as different valleys can be distinguished.
Based on the topological argument, these DW channels are insensitive to the detailed DW profile, e.g., the orientation, and
as long as the DW is smooth at the lattice scale and the intervalley scattering way below the Fermi surface is also suppressed.
As one can observe in Fig.~\ref{fig1}(e), the DW mode typically spans more than few tens of lattice sites even for a narrow wall and hence any scattering involving transfer of a sizable momentum becomes virtually impossible.

{\color{blue}{\em Field control of valley filter.}}---
Now we examine a more realistic system where spin is included and demonstrate the versatile operating modes of the valley filter
which can be controlled by external electric and magnetic fields. Consider a tight-binding model $\mathcal{H}_0$ of a honeycomb lattice with the usual nearest neighbor hopping, as the $t$ term in Eq.~(\ref{toyH}).
Again, the QVH domain can be realized by a $\mathcal{P}$-breaking staggered sublattice potential $\mathcal{H}_v$,
as the $\lambda_v$ term in Eq.~(\ref{toyH}).
Physically, such a staggered potential can arise from the interaction with a substrate~\cite{zhou2007}
or from the occupation of the two sublattices by different atoms or atomic orbitals.~\cite{xiao2012,li2013}
Of particular interest is the case in which the two sublattices are slightly shifted
with respect to each other out-of-plane forming a buckled structure.
Such buckling has been found in 2D materials such as silicene,~\cite{caha2009} germanene,~\cite{caha2009}
and quite a few chemically functionalized materials.~\cite{jun2010,xu2013,ccliu2014,song2014}
In this class of materials $\mathcal{H}_v$ can be induced and tuned by a perpendicular electric field.

Regarding the QAH domain, it can be engineered by a Rashba spin-orbit coupling
$\mathcal{H}_R=it_R\sum_{\langle i,j\rangle,\alpha\beta}c_{i\alpha}^\dagger(\bm s^{\alpha\beta}\times\bm d_{ij})_z c_{j\beta}$
combined with an exchange coupling $\mathcal{H}_M=M\sum_{i,\alpha\beta}c_{i\alpha}^\dagger s_z^{\alpha\beta}c_{i\beta}$,
where $\bm d_{ij}$ is a unit vector pointing from site $j$ to site $i$ and $\bm s$ are the spin Pauli matrices.
The $\mathcal{T}$-breaking exchange coupling lifts the spin degeneracy producing a band inversion,
whereas the $\mathcal{P}$-breaking Rashba term hybridizes the inverted bands yielding an energy gap.
This QAH model has been proposed in the context of graphene.~\cite{qiao2010}
The Rashba and the exchange couplings may be produced simultaneously by ferromagnetic adsorbed atoms
or in proximity to a (anti)ferromagnetic substrate.~\cite{qiao2010,ding2011,bala2013}

\begin{figure}
\scalebox{0.42}{\includegraphics*{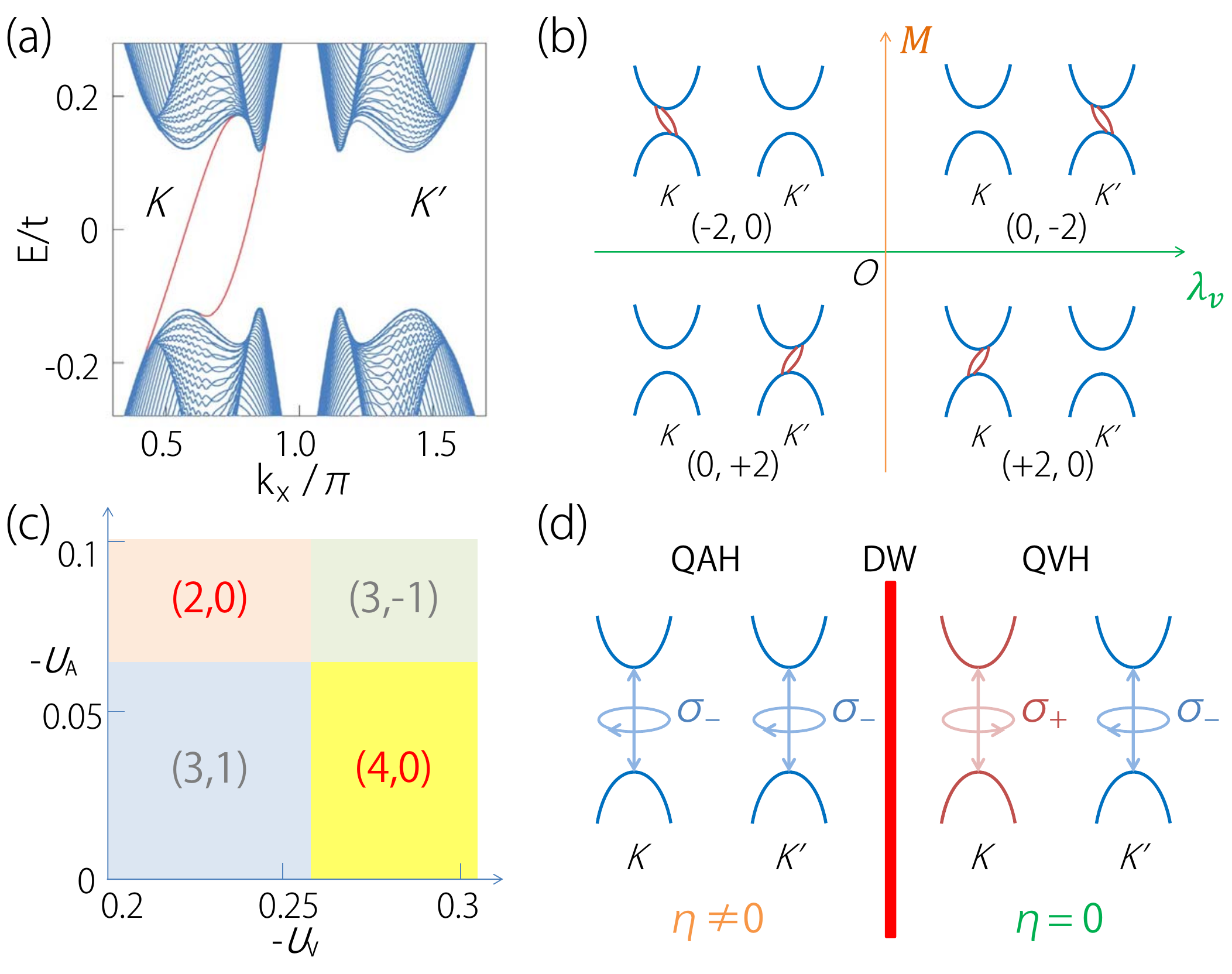}}
\caption{(a) Energy spectrum of the monolayer model featuring two valley-filtered chiral DW channels (in red color).~\cite{param_fig2}
(b) $(\nu_K,\nu_{K'})$ as a function of the exchange coupling $M$ and the sublattice staggered potential $\lambda_v$ in the monolayer model.
(c) $(\nu_K,\nu_{K'})$ in the bilayer model, tuned by the interlayer potential differences $U_V$ and $U_A$ on the two sides of the DW.~\cite{param_fig2}
(d) Optical circular dichroism (with light frequency of the bulk gap) exhibiting contrasting feature across the DW.}
\label{fig2}
\end{figure}

A straightforward calculation yields that the valley-contrasting topological charge is $\tau_z\text{sgn}(\lambda_v)$
for the QVH state with $\mathcal{H}_\text{QVH}=\mathcal{H}_0+\mathcal{H}_v$ and $\text{sgn}(M)$
for the QAH state with $\mathcal{H}_\text{QAH}=\mathcal{H}_0+\mathcal{H}_R+\mathcal{H}_M$, respectively.
Therefore, at a DW between the two states, the number of chiral gapless DW channels at valley $\tau_z$ must be
\begin{equation}\label{slnu}
\nu=\tau_z\text{sgn}(\lambda_v)-\text{sgn}(M),
\end{equation}
based on the bulk-boundary correspondence as we discussed in the previous section.
Again the important consequence is that the chiral gapless DW channels must appear and only appear in a single valley.
As an example, when $\lambda_v>0$ and $M<0$, $\nu=2$ for valley $K$ and $\nu=0$ for valley $K'$.
This implies there are two DW channels propagating in $+\hat{x}$ direction at valley $K$ and no state at valley $K'$,
which is confirmed by our numerical calculation and shown in Fig.~\ref{fig2}(a).
These DW states can also be solved in a low-energy continuum model, similar to what we have done for the toy model.

It follows from Eq.~(\ref{slnu}) that both the valley index of the DW channels and their propagating directions (chirality)
are determined by the signs of $\lambda_v$ and $M$, which may be manipulated by tuning external electric and magnetic fields.
In Fig.~\ref{fig2}(b), we schematically show how the characters of DW channels vary as the signs of $\lambda_v$ and $M$ are switched.
The two axes ($\lambda_v=0$ and $M=0$) correspond to topological phase boundaries at which the bulk gap vanishes.
Eq.~(\ref{slnu}) and Fig.~\ref{fig2}(b) are suggestive of a versatile control of the operating modes for the valley filter.
As an example, consider an initial mode with $\lambda_v>0$ and $M<0$.
Flipping only the sign of $\lambda_v$, the valley index becomes opposite whereas the chirality remains the same.
Inverting both the signs of $\lambda_v$ and $M$, the valley index maintains whereas the chirality is reversed.
Both valley and chirality modes can be switched simultaneously by a sign reversal of $M$.
All these characters can also be easily argued from the symmetry point of view.

All the above discussions can be directly extended to a bilayer system that may
exhibit a richer topological phase diagram due to the additional layer degree of freedom.
We consider an AB-stacked bilayer graphene model following Ref.~\onlinecite{qiao2013}.
Each layer is described by the same Hamiltonian $\mathcal{H}_0+\mathcal{H}_R+\mathcal{H}_M$ as discussed above.
The two layers are coupled by the interlayer hopping, $\mathcal{H}_t=t_\bot\sum_{\langle i\in (t,A),j\in(b,B)\rangle}c_i^\dagger c_j$,
between the A site of top layer and the B site of bottom layer that are the nearest interlayer neighbors.
Through dual gating, it is convenient to introduce an interlayer potential difference $\mathcal{H}_U=U\sum_{i}\zeta_i c_i^\dagger c_i$
with $\zeta_i=\pm$ denoting the top and bottom layers.
$\mathcal{H}_U$ plays a similar role here to that of the staggered sublattice potential $\mathcal{H}_v$ in the buckled monolayer system.
Depending on the competition between the exchange coupling $\mathcal{H}_M$ and other potentials,
various distinct QVH and QAH states can be realized.~\cite{qiao2013,PanH2,Pan2014}

Compared with the monolayer case, of particular interest in the bilayer case is that its valley-projected topological charge
can have a magnitude of either one or two, depending on the model parameters.~\cite{qiao2013}
As a consequence, at the DW between the bilayer QAH and QVH states,
there exist two types of perfect valley filters with different number of chiral DW channels:
for one type it has two perfect valley-filtered channels, whereas for the other type it has four.
In Fig.~\ref{fig2}(c), we plot $(\nu_K,\nu_{K'})$, the number of chiral channels in each valley,
as a function of $U_A$ and $U_V$ with other parameters fixed.
Here we have assumed that the interlayer potential differences on the two sides of the DW, denoted by $U_A$ ($U_V$) for the QAH (QVH) domain,
can be controlled independently. One notes that in the chosen parameter range, there are four regions of $(\nu_K,\nu_{K'})$,
of which $(2,0)$ and $(4,0)$ are perfect valley filters whereas the other two have DW channels in both valleys.
Nonetheless, this indicates that by controlling the interlayer potentials, it is possible to tune the conductance of a perfect valley filter,
e.g., from $2e^2/h$ to $4e^2/h$ when the phase switches from $(2,0)$ to $(4,0)$.

{\color{blue}{\em Optical circular dichroism.}}---
Conventionally, gapless DW modes can be probed by transport or scanning tunneling spectroscopy.
Since a perfect valley filter requires the breaking of $\mathcal{T}$, optical circular dichroism may also be employed to detect
the presence, the location, and the characters of the topological DW bearing a perfect valley filter.
We sketch the circular dichroism in Fig.~\ref{fig2}(d) for our model
and illustrate the physics below using the standard argument of Fermi's golden rule.
The coupling strength of an interband optical transition with $\sigma_\pm$ circularly-polarized light is given by
$\mathcal{M}_\pm(\bm k)=\mathcal{M}_x\pm i\mathcal{M}_y$,
where $\mathcal{M}_\mu=\langle u^c_{\bm k}|\nabla_{k_\mu} H|u^v_{\bm k}\rangle$ is the interband matrix element.
For a valley-projected chiral two-band system~\cite{zhan2011} with a gap $2\Delta$, e.g., gapped monolayer or bilayer graphene,
the absorbance~\cite{yao2008,tru2011} of the $\sigma_\pm$ light with frequency $\omega$ has a universal form
$\propto(1\pm 2\tau_z{\Delta}/{\omega})^2\Theta(\omega-2\Delta)$.
When the optical frequency is close to the band gap,
at valley $K$ ($K'$) the interband transition occurs only for $\sigma_+$ ($\sigma_-$) circular polarization.
For the QVH state, the net circular dichroism
$\eta= (|\mathcal{M}_+|^2-|\mathcal{M}_-|^2)/(|\mathcal{M}_+|^2+|\mathcal{M}_-|^2)$ vanishes, as required by $\mathcal{T}$.
In contrast, for the QAH state breaking $\mathcal{T}$, there is a strong circular dichroism.
Therefore, such a difference between the two domains in optical absorbance or photoluminescence
can be used to probe the presence and the location of the valley filter.
Since $|\mathcal{M}_\pm|^2$ also depend on the sign of $\Delta$,
the circular dichroism can also probe to the character of the valley filter that is sensitive to the sign of $\Delta$.

{\color{blue}{\em Discussion.}}---
We have proposed a general scheme for creating perfect valley-filtered channels at the DW between two topologically-distinct insulating regions.
The ease of local field control of each filter will allow the construction of more complex valleytronic devices
using multiple DWs or a percolating DW network.
For example, a valley valve can be constructed by combining two valley filters in series.
Whether a current can pass through the valve depends on whether the two filters have the same valley index and the same chirality,
which can be controlled by local electric and/or magnetic fields.
Although it is not our focus here to investigate specific materials for its realization,
we do note that it is promising to obtain the desired topological DWs in 2D materials with hexagonal lattice symmetry such as single or multilayer graphene, silicene, and chemically functionalized monolayer materials, through controlled doping, gating, and substrate effects.
Our scheme can be generalized to materials with more than two valleys,~\cite{jifeng}
and a perfect valley filter can also be created whenever across a DW the valley-projected topological charge only changes at one designed valley.

{\indent{\em Acknowledgments.}}---The authors thank D.L. Deng for helpful discussions. H.P. is supported by the NSFC under Grant No. 11174022, F.Z. is supported by UT Dallas research enhancement funds, and S.A.Y. is supported by SUTD-SRG-EPD2013062.

\bibliographystyle{apsrev4-1}

\begin{thebibliography}{100}
\bibitem{param_fig1} We have chosen $t_2=0.02$ ($0$) and $\lambda_v=0$ ($0.5$) for the QAH (QVH) state.
\bibitem{param_fig2} In Fig.2(a), we have chosen $t_R=M=0$ and $\lambda_v=0.5$ for the QVH state,
                     and $t_R=0.1$, $M=-0.5$, and $\lambda_v=0$ for the QAH state.
                     In Fig.2(c), we have chosen $t_R=0.1$ and $M=0$ for QVH state,
                     $t_R=0.02$ and $M=-0.15$ for QAH state, and $t_\bot=0.15$ for both cases.
\end{thebibliography}

\end{document}